\documentclass{emulateapj}
\shorttitle{Evolution in average SN Ia properties with redshift}
\shortauthors{Howell}

\begin{document}
\title{Predicted and observed evolution in the mean properties of Type~Ia supernovae
  with redshift}
\author{D. Andrew Howell, Mark Sullivan, Alex Conley, Ray Carlberg}
\affil{Department of Astronomy and Astrophysics, University of
  Toronto, 50 St. George St., Rm. 101, Toronto, ON, Canada M5S 3H4}
\begin{abstract}
  Recent studies indicate that Type Ia supernovae (SNe Ia) consist of
  two groups --- a ``prompt'' component whose rates are proportional
  to the host galaxy star formation rate, whose members have broader
  lightcurves and are intrinsically more luminous, and a ``delayed''
  component whose members take several Gyr to explode, have narrower
  lightcurves, and are intrinsically fainter.  As cosmic star
  formation density increases with redshift, the prompt component
  should begin to dominate.  We use a two-component model to predict
  that the average lightcurve width should increase by 6\% from
  $z=0-1.5$.  Using data from various searches we find an
  8.1\%$\pm$2.7\% increase in average lightcurve width for
  non-subluminous SNe Ia from $z=0.03 - 1.12$, corresponding to an
  increase in the average intrinsic luminosity of 12\%.  To test
  whether there is any bias after supernovae are corrected
  for lightcurve shape we use published data to mimic the effect of
  population evolution and find no significant difference in the
  measured dark energy equation of state parameter, $w$.  However,
  future measurements of time-variable $w$ will require
  standardization of SN Ia magnitudes to 2\% up to $z=1.7$, and it is
  not yet possible to assess whether lightcurve shape correction works
  at this level of precision.  Another concern at $z=1.5$ is the
  expected order of magnitude increase in the number of SNe Ia that
  cannot be calibrated by current methods.
\end{abstract}

\keywords{cosmology: observations --- cosmology: cosmological
  parameters --- supernovae: general --- surveys}

\section{Introduction}\label{intro}
Type Ia supernovae (SNe Ia) are the most important standardized
candles for cosmology, and have been used to discover dark energy and
the accelerating universe \citep{1998AJ....116.1009R,
  1999ApJ...517..565P}.  This was facilitated by the realization that
supernovae with broader lightcurves are intrinsically brighter, while
those with narrow lightcurves are dimmer \citep{1993ApJ...413L.105P}.
Various schemes exist to correct SN Ia luminosities based on their
lightcurve shape \citep[e.g.][]{1996ApJ...473...88R,
  2003ApJ...594....1T, 2006ApJ...647..501P,
  2007A&A...466...11G,2007ApJ...659..122J} --- here we use the
``stretch'' method, in which the time axis of a template lightcurve is
multiplied by a scale factor $s$ to fit the data
\citep{1997ApJ...483..565P}.

Some properties of SNe Ia have been found to correlate with
environment --- brighter supernovae with broader lightcurves (high
$s$) tend to occur in late-type spiral galaxies
\citep{1995AJ....109....1H}, while dimmer, fast declining (low $s$)
supernovae are preferentially located in an older stellar population,
leading to the conclusion that the age of the progenitor system is a
key variable affecting SN Ia properties \citep{2001ApJ...554L.193H}.
The fact that supernovae occur at a much higher rate in late type
galaxies, and that the SN Ia rate is proportional to the core-collapse
rate \citep{2005A&A...433..807M} is another indication that age plays
an essential role.  Following this previous work,
\citet{2005ApJ...629L..85S} model SNe Ia as consisting of two
populations --- a ``prompt'' component whose rate is proportional to
the star formation rate of the host galaxy, and a second ``delayed''
component whose rate is proportional to the stellar mass of the
galaxy.  \citet[hereafter S06]{2006ApJ...648..868S} tie all of these
results together using data from the Supernova Legacy Survey (SNLS),
finding that slow declining, brighter SNe Ia come from a young
population and have a rate proportional to star formation on a 0.5 Gyr
timescale, while dimmer, faster declining supernovae come from a much
older population with a rate proportional to the mass of the host
galaxy.

S06 and \citet{2006MNRAS.370..773M} predict that the
SNe Ia whose rates are proportional to star formation will start to
dominate the total sample of SNe as cosmic star formation
increases with redshift.  Since these SNe are intrinsically
brighter, the mean luminosity of the population should increase with
redshift.  Here we use the two-component SN Ia model of 
\citet{2005ApJ...629L..85S} and the stretch distribution for
each component from SNLS data (S06) to quantify the expected magnitude
of this effect.  We then compare the predicted evolution in lightcurve
stretch to SN distributions from the SNLS and the Higher-z SN Search
\citep{2007ApJ...659...98R}.

\section{Predicting Evolution}\label{method}
\citet{2005ApJ...629L..85S} parameterize the supernova rate in a
galaxy as $$\rm{SNR(t)}=\it A M(t) + B \dot{M}(t),$$
where $M(t)$ is
the total stellar mass in the galaxy, $\dot{M}$ is the star formation
rate, and A and B are constants.  These authors 
use the supernova
rates in galaxies of different morphologies and colors to derive
values for $A$ and $B$.  S06 use an alternate method --- they fit
galaxy models to SNLS $u^*g\arcmin r\arcmin i\arcmin z\arcmin$ host
galaxy photometry to derive masses and star formation rates.  Then,
using the cosmic star formation history of
\citet{2006ApJ...651..142H}, S06 predict the rate of SNe from each
component versus redshift in their Fig. 10.  Note that here we adopt
the same definition of $\dot{M}(t)$ as S06 --- it is the mass turned
into stars and does not include mass loss from supernovae.

\begin{figure}
\epsscale{0.7}
\plotone{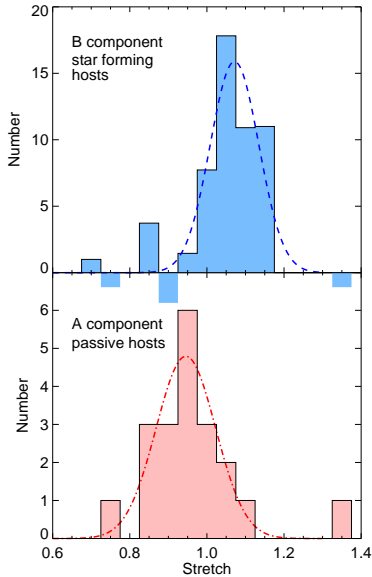}
\caption{Stretch distributions of SNe from each component of the two
  component model.  {\em Top:} 
Prompt ($B$-component) SNe Ia.  {\em Bottom:} Delayed ($A$-component)
SNe Ia.  The distributions were derived from those in S06, as
described in the text.  Best-fit Gaussians are shown.  The
$A$-component Gaussian is centered at $s=0.945$, with $\sigma=0.077$.
The $B$-component Gaussian is centered at $1.071$ with $\sigma=
0.063$.}
\label{abdist}
\end{figure}

The prompt and delayed SN Ia components have different stretch
distributions (S06).  
To determine the stretch of each SN, here we use
the SNe from S06, though we fit a new lightcurve template to the data
using the SiFTO method \citep{2007arXiv0705.0367C,2007conley}.
Because stretches are always defined relative to the $s=1$ template
lightcurve, stretch values should only compared within a publication.
However, the stretches derived here are approximately 4\% larger than
those in \citet{2006A&A...447...31A}, largely due to the use of a
narrower $s=1$ template.

All SNe from passive galaxies (i.e. those with no measurable star
formation rate) were assigned to the $A$ component.  Star forming
galaxies have SNe Ia from both components, so the $A$ distribution
from passive galaxies was scaled by mass and subtracted from the
distribution of SNe Ia from star forming galaxies, leaving the $B$
distribution (as in S06).  The resulting distributions, and Gaussian
fits are shown in Figure~\ref{abdist}.  Note that to conserve the
total number of SNe one should add the SNe subtracted from the $B$
distribution back to the $A$ distribution -- this is unnecessary for
our purposes because the relative heights of the gaussians are
normalized as a function of redshift in the next step.

To estimate the expected stretch evolution with redshift, we take the
observed $A$ and $B$ distributions and scale them to the predicted
relative values with redshift from Fig. 10 of S06.  Increasing cosmic
star formation with redshift produces a larger fraction of SNe from
the prompt component.  Stellar mass as a function of redshift is
determined by integrating the star formation history from the earliest
times, so the total stellar mass, and the number of SNe from the A
component, decreases with increasing redshift.  The net result is that
in the $A+B$ model the mean stretch increases from 0.98 at $z=0$ to
1.04 at $z=1.5$.

One caveat is that in the $A+B$ model there is no time dependence for
the $A$ component.  SNe Ia from 10 Gyr old progenitors are just as
likely as SNe Ia from 3 Gyr old progenitors.  If 10 Gyr-old SNe Ia are
actually more rare, the $A+B$ model will overpredict the number of
$A$-component SNe at $z=0$, as they result from stars formed during
the high star formation rate in the early universe (see discussion in
S06).  As an alternative to the $A+B$ model we tested the two
component SN Ia delay time distribution from
\citet{2006MNRAS.370..773M}, which has an exponential decrease in
supernova probability from the delayed component with time.  The
drawback of this model is that the probability distribution is
somewhat arbitrary.  Also, rather than the 50-50 split between prompt
and delayed SNe chosen by \citet{2006MNRAS.370..773M}, here we scale
each component by the $A$ and $B$ values measured by S06.  This gives
similar results to the $A+B$ model, predicting a shift in mean stretch
from 0.98 to 1.02 from $z=0-1.5$.  

\section{Comparison to Observations}
\begin{figure}
\plotone{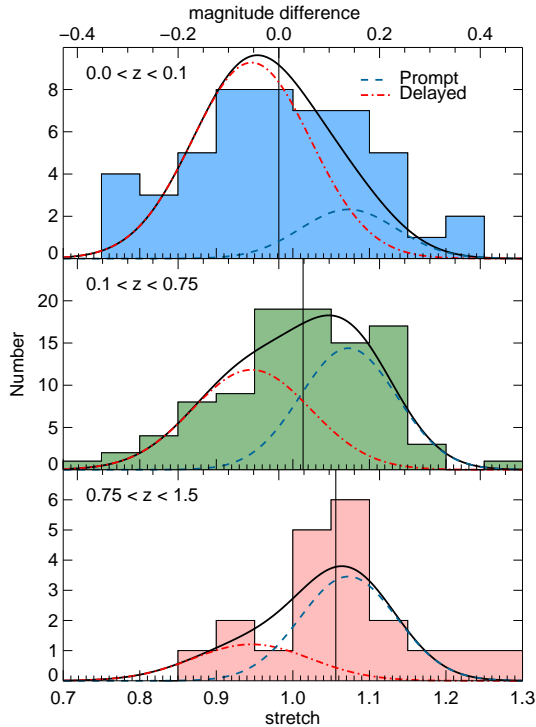}
\caption{Actual stretch distributions compared to predictions from the
  $A+B$ model.  In each case the $A+B$ model evaluated at the median
  redshift of the distribution is shown.  {\em Top:} SNe Ia from
  $z=0-0.1$ (median $z=0.026$; N=50).  {\em Middle:} SNLS SNe Ia from
  $z=0.1-0.75$ (median $z=0.55$; N=99).  {\em
    Bottom:} \citet{2007ApJ...659...98R} SNe Ia from $z=0.75-1.5$
  (median $z=1.12$; N=20), with
  $s$ errors $< 0.2$ (requiring $s$ error $\leq 0.1$ reduces the
  sample to 16 and increases the average stretch from 1.06 to 1.07).  
  The vertical line gives the mean stretch for each distribution.
  The top axis converts a stretch difference into a magnitude
  difference using $\alpha=1.5$.}
\label{gaussreal}
\end{figure}

In Figure~\ref{gaussreal} we compare the predicted stretch
distributions from the $A+B$ model to the observed stretch
distributions in three redshift bins from the low redshift data used
by \citet{2006A&A...447...31A}, the SNLS data in S06, and the data of
the higher-z supernova search \citep{2007ApJ...659...98R}.  All
lightcurves have been refit here using the same method.  We also
tested the data against the modified \citet{2006MNRAS.370..773M}
model, but we did not find it to be a better predictor of SN evolution
with redshift (Table~\ref{table}). 

\begin{deluxetable}{ccccc}
\tablewidth{0pt} 
\tablecolumns{5}
\tablecaption{$\chi^2$ and KS test: data and models\label{table}} 
\tablehead{
\colhead{} & \multicolumn{2}{c}{$\chi^2$} & \multicolumn{2}{c}{KS} \\
\colhead{z} & \colhead{A+B} &\colhead{MM} & \colhead{A+B} & \colhead{MM} 
} 
\startdata
0-0.1 & 0.81 & 0.63 & 15\% & 39\% \\
0.1-0.75 & 0.64 & 0.83 & 30\% & 28\% \\
0.75-1.5 & 0.60 & 0.84 & 52\% & 35\% \\
\enddata 
  
\tablecomments{Cols. 2-3: the $\chi^2$ per
    degree of freedom between the data and the predictions of the A+B
    and modified Mannucci (MM) models.  Cols. 4-5: KS-test probability that
    the data is drawn from each model.  Bins with zero counts were
    assigned an error of 1.15, possibly underestimating the $\chi^2$.
    The KS test is also imperfect because probabilities were
    derived for a single Gaussian, not the sum of two Gaussians as
    used here.  In both cases the two model components were fixed by
    the A and B numbers in S06.}
\end{deluxetable}

Each survey has different selection effects -- the most serious for
the current study is Malmquist bias, the tendency to discover only the
brightest members of a group near the detection limit of a
magnitude-limited survey.  To minimize the effect, for each of the
high redshift searches we only consider supernovae from a reduced
volume so that none of the supernovae used are near the magnitude
limit.  The SNLS regularly discovers SNe Ia out to $z>1$, but here we
use only the subset with $z \leq 0.75$, where Malmquist bias is
minimal \citep{2006A&A...447...31A}.  Similarly, we only use
\citet{2007ApJ...659...98R} SNe with $z<1.5$, where the authors report
their sample is complete \citep{2004ApJ...613..200S}.  Lowering the
redshift cuttoff to $z=1.2$ does not change the average stretch for
the highest-z sample, but it reduces the sample size from 20 to 13,
and thus decreases the significance of the results.

As an additional protection against selection bias, we only consider
SNe with $s \geq 0.7$.  SNe Ia with $s<0.7$ are both dim and
spectoscopically peculiar, like SN 1991bg \citep{1992AJ....104.1543F},
and have not yet been detected at $z>0.2$, probably because of a
combination of Malmquist bias and spectroscopic selection bias
\citep{2001ApJ...554L.193H,2005ApJ...634.1190H} -- as redshift
increases, and the angular size of the host galaxy decreases, and it
becomes more and more difficult to spectroscopically identify such
faint supernovae when blended with their bright, often elliptical,
hosts.  This cut removes 3 SNe Ia from the low-z sample [other low-z
SNe are already removed because we only consider Hubble-flow SNe Ia,
with $z>0.015$, to be consistent with \citet{2006A&A...447...31A}].

In all cases we use only SNe Ia with at least 4 lightcurve points, and
at least one detection before 10 rest-frame days after maximum light in the
$B$-band, so that stretch is accurately measured.

Figure~\ref{gaussreal} shows that the average observed stretch
increases with redshift, from $s=0.98\pm 0.02$ at a median redshift of
$z=0.03$, to $s=1.02\pm 0.01$ at $z=0.56$, and $s=1.06\pm 0.02$ at
$z=1.12$.  Simultaneously the percentage of SNe Ia with $s<0.9$
decreases from 24\% to 15\% to 1.4\%.  The KS test gives a 2\%
probability that the lowest and highest redshift sample are drawn from
the same distribution.  The predicted distributions from the $A+B$
model are overplotted.  The observed trends match the predictions of
the empirically-based models --- with increasing redshift fewer
low-stretch SNe Ia are observed, and the mean SN Ia stretch increases.
We find the same result when this analysis is repeated with the SALT
\citep{2005A&A...443..781G} and SALT2 \citep{2007A&A...466...11G}
lightcurve fitters.    These results are also consistent with the
findings of \citet{2006A&A...447...31A}, that the low-z sample had an
average stretch 97\% that of SNLS SNe.

\citet{2006A&A...447...31A} estimate distances from SNe Ia using 
$$\mu_B = m^*_B -M + \alpha (s-1) - \beta c,$$ where $\mu_B$ is the 
distance modulus, $m^*_B$ is the peak $B$-band magnitude, $c$ is a
color, and $M$, $\alpha$ and $\beta$ are parameters fit by minimizing
residuals on the Hubble diagram.
\citet{2006A&A...447...31A} found $\alpha = 1.52$, so a
drift in average stretch of $0.08 \pm 0.026$ from $z=0.03$ to $z=1.12$
results in a 12\% drift in average intrinsic SN Ia luminosity
over this redshift range.

\section{Effect on Cosmological Studies}
Evolution in the SN population will not necessarily bias cosmological
studies, since SNe are only used in this way after correction for
lightcurve shape.  However, we can no longer assume that any
deficiencies in lightcurve width correction schemes will average out
under the assumption the distribution of SNe is similar over 
all redshifts.  If there is a systematic residual between
low stretch and high stretch SNe when they are stretch
corrected, this could cause a bias in the determination of
cosmological parameters as the population evolves.

\begin{figure}
\plotone{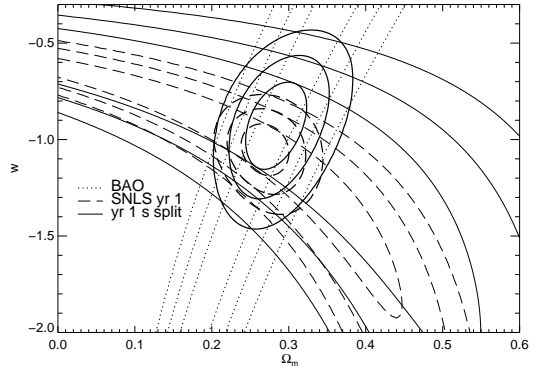}
\caption{Cosmological fits done with the \citet{2006A&A...447...31A}
  sample (dashed), and with SNe Ia drawn from the same sample, but
  using only $s<1$ SNe Ia at $z<0.4$ and $s\geq 1$ SNe Ia at 
  $z\geq 0.4$ (solid).  Combining with Baryon Acoustic Oscillation results
  \citep{2005ApJ...633..560E} produces the thick lines.
  The results are consistent at the one sigma level.}
\label{lowhighs}
\end{figure}

To test an extreme case of evolution, we fit the equation of state
parameter for dark energy, $w$ and the matter density, $\Omega_M$
using the data from \citet{2006A&A...447...31A}, (dashed lines in
Fig.~\ref{lowhighs}) and compared it to a fit using the same data, but
retaining only $s<1$ SNe Ia at $z<0.4$ and only $s\geq 1$ SNe Ia at
$z\geq 0.4$ (solid lines).  The strongly evolving subset gives estimates
for $w$ and $\Omega_M$ consistent with the full set, although
the errors are larger because there are fewer SNe in the subset.

One worry with an evolving population is that stretch-magnitude or
color-magnitude relations may evolve, i.e. $\alpha$ or $\beta$ derived
at one redshift may not be appropriate at another.  The values derived
here for the strong evolution subset and the full set are consistent,
but again a strong test awaits a larger data set.

Measuring changes in $w$ with time will require much stricter control
of potential evolutionary effects.  Possible biases depend critically on the
exact nature of the evolution, the experimental design, the cosmology, and how the time 
variable component of $w$ is parameterized --- e.g. $w\prime $, $w_1$,
or $w_a$
\citep{2002PhRvD..65j3512W,2004MNRAS.347..909K,2006APh....26..102L}.
However, a good rule of thumb is that to keep systematic errors 
significantly below statistical errors for a mission such as SNAP, the
corrected magnitudes of SNe Ia should not drift by more than 0.02 mag
up to $z=1.7$ \citep{2004MNRAS.347..909K}.  Unfortunately there are
not enough well measured SNe Ia in the literature to determine whether
or not stretch correction works to this level of precision.  
After stretch correction, the rms scatter of SNe Ia around the Hubble 
line is $\sim 0.2$ mag in the best cases \citep{2007A&A...466...11G}.
Therefore $\sim 100$ SNe Ia are required in each of several stretch or
redshift bins to determine whether biases remain at the 0.02 level 
after correction.  Such precise tests will soon be possible by
combining data from large surveys underway at low and high redshift.

\section{Discussion}
We have shown that there is some evolution in the average lightcurve
width, and thus intrinsic luminosity of SNe Ia from $z=0$ to $z=1.5$,
although significant evolution is found only over a large redshift
baseline.  This evolution is consistent with our predictions from the
$A+B$ model -- as star formation increases with redshift, the
broader-lightcurve SNe Ia associated with a young stellar population
make up an increasingly larger fraction of SNe Ia.

Though we have taken steps to minimize the effects of Malmquist bias, 
it is possible that residual 
effects play some role in increasing average stretch with redshift.
However, the net effect on cosmology studies is the same no matter 
the underlying cause.  In either case, there is increased pressure on
the light curve shape calibration method to correct for the evolution in 
SN Ia properties with redshift.

\citet{1999ApJ...517..565P} found that SNe Ia still give evidence for
an accelerating universe even if SNe Ia are not corrected for stretch.
This was possible because the difference in a universe with dark
energy and one with $\Omega_M=1$ is large --- 0.25 magnitudes at
$z=0.7$, whereas the population evolution seen here implies that average SN
Ia magnitudes should increase by 0.07 mag over the same redshift
range.  However, discriminating between dark energy models requires
much more precise control of SN Ia magnitudes over a larger redshift
baseline. 

Most theoretical studies addressing possible SN Ia evolution have
focused on metallicity.  Although theorists have proposed various
mechanisms that could conceivably alter the properties of SNe Ia as
the average metallicity changes with cosmic time
\citep{1998ApJ...495..617H,2000ApJ...530..966L, 2001ApJ...557..279D,
  2003ApJ...590L..83T} there is no consensus
regarding which effects are important or even the sign of these
effects.  Observational studies have found no evidence that
metallicity affects the properties of SNe Ia
\citep{2000AJ....120.1479H,2000ApJ...542..588I,2005ApJ...634..210G,2007ellis}.
Instead, it is more likely that age differences between the two
populations (and thus almost certainly the mass of the secondary star)
play a role in the evolution of the observed stretch distribution with
redshift \citep{2001ApJ...554L.193H}.

Prompt SNe Ia are thought to be brighter because they produce more $^{56}$Ni.
If the Chandrasekhar-mass model describes most SNe Ia, they must then
produce less intermediate mass elements, assuming that the amount of
unburned material is negligible in normal SNe Ia.  We therefore
predict that high redshift SNe Ia will have less Ca and Si.  
This is confirmed by the most intensive study of
high redshift SN Ia spectra \citep{2007ellis}.

One concern raised by these findings is that pathological SNe Ia such
as SN 2001ay \citep{howell2004}, SN 2002cx
\citep{2003PASP..115..453L}, SN 2002ic \citep{2003Natur.424..651H},
and SNLS-03D3bb \citep{2006Natur.443..308H},
which do not obey typical lightcurve shape correction schemes, are
associated with star formation.  Since star formation density
increases by a factor of
ten from $z=0$ to $z=1.5$ \citep{2006ApJ...651..142H},
at high redshift these pathological supernovae will be an order of
magnitude more common.  Thus the conventional wisdom that all high
redshift supernovae will have counterparts at low redshift 
\citep{2001astro.ph..9070B} only holds if sample sizes are much larger
than those currently used for cosmology.  Only 20 SNe~Ia have
published lightcurves at $z>1$
\citep{2007ApJ...659...98R,2006A&A...447...31A}, and only $\sim $50
SNe Ia at $z<0.1$ have sufficient data to be cosmologically useful.
Fortunately, thus far it has been possible to identify these outliers
so that they do not affect cosmological analyses, but future studies
requiring increased precision must be vigilant of the effects of
an evolving SN Ia population.

\acknowledgements
We thank Richard Ellis, Peter Garnavich, Julien Guy, Eric Linder, Peter Nugent, and an
anonymous referee for helpful comments and the Canadian NSERC for support.

\bibliographystyle{apj}

\begin{thebibliography}{41}
\expandafter\ifx\csname natexlab\endcsname\relax\def\natexlab#1{#1}\fi

\bibitem[{{Astier} {et~al.}(2006)}]{2006A&A...447...31A} 
{Astier}, P. {et~al.} 2006, \aap, 447, 31 

\bibitem[{{Branch} {et~al.}(2001)}]{2001astro.ph..9070B} 
{Branch}, D., {Perlmutter}, S., {Baron}, E., \& {Nugent}, P.  2001, ArXiv  
Astrophysics e-prints 

\bibitem[{{Conley} {et~al.}(2007{\natexlab{a}})}]{2007arXiv0705.0367C} 
{Conley}, A., {Carlberg}, R.~G., {Guy}, J., {Howell}, D.~A., {Jha}, S.,  
{Riess}, A.~G., \& {Sullivan}, M.  2007{\natexlab{a}}, ArXiv e-prints, 705 

\bibitem[{{Conley} {et~al.}(2007{\natexlab{b}})}]{2007conley} 
{Conley}, A. 2007{\natexlab{b}}, in preparation 

\bibitem[{{Dom{\' i}nguez}, {H{\" o}flich}, \&  
{Straniero}(2001)}]{2001ApJ...557..279D} 
{Dom{\' i}nguez}, I., {H{\" o}flich}, P., \& {Straniero}, O.  2001, \apj, 
557,  279 

\bibitem[{{Eisenstein} {et~al.}(2005)}]{2005ApJ...633..560E} 
{Eisenstein}, D.~J. {et~al.} 2005, \apj, 633, 560 

\bibitem[{{Ellis} {et~al.}(2007)}]{2007ellis} 
{Ellis}, R. 2007, in preparation 

\bibitem[{{Filippenko} {et~al.}(1992)}]{1992AJ....104.1543F} 
{Filippenko}, A.~V. {et~al.} 1992, \aj,  104, 1543 

\bibitem[{{Gallagher} {et~al.}(2005)}]{2005ApJ...634..210G} 
{Gallagher}, J.~S., {Garnavich}, P.~M., {Berlind}, P., {Challis}, P., 
{Jha},  S., \& {Kirshner}, R.~P.  2005, \apj, 634, 210 

\bibitem[{{Guy} {et~al.}(2007)}]{2007A&A...466...11G} 
{Guy}, J. {et~al.} 2007, \aap, 466, 11 

\bibitem[{{Guy} {et~al.}(2005)}]{2005A&A...443..781G} 
{Guy}, J., {Astier}, P., {Nobili}, S., {Regnault}, N., \& {Pain}, R.  2005, 
 \aap, 443, 781 

\bibitem[{{Hamuy} {et~al.}(1995)}]{1995AJ....109....1H} 
{Hamuy}, M., {Phillips}, M.~M., {Maza}, J., {Suntzeff}, N.~B., {Schommer},  
R.~A., \& {Aviles}, R.  1995, \aj, 109, 1 

\bibitem[{{Hamuy} {et~al.}(2003)}]{2003Natur.424..651H} 
{Hamuy}, M. {et~al.} 2003, \nat, 424, 651 

\bibitem[{{Hamuy} {et~al.}(2000)}]{2000AJ....120.1479H} 
{Hamuy}, M., {Trager}, S.~C., {Pinto}, P.~A., {Phillips}, M.~M., 
{Schommer},  R.~A., {Ivanov}, V., \& {Suntzeff}, N.~B.  2000, \aj, 120, 
1479 

\bibitem[{{H\"oflich}, {Wheeler}, \&  
{Thielemann}(1998)}]{1998ApJ...495..617H} 
{H\"oflich}, P., {Wheeler}, J.~C., \& {Thielemann}, F.~K.  1998, \apj, 495, 
617 

\bibitem[{{Hopkins} \& {Beacom}(2006)}]{2006ApJ...651..142H} 
{Hopkins}, A.~M. \& {Beacom}, J.~F.  2006, \apj, 651, 142 

\bibitem[{{Howell}(2001)}]{2001ApJ...554L.193H} 
{Howell}, D.~A.  2001, \apjl, 554, L193 

\bibitem[{{Howell} \& {Nugent}(2004)}]{howell2004} 
{Howell}, D.~A. \& {Nugent}, P.  2004, in Cosmic Explosions in Three  
Dimensions, 151 

\bibitem[{{Howell} {et~al.}(2006)}]{2006Natur.443..308H} 
{Howell}, D.~A. {et~al.} 2006, \nat, 443, 308 

\bibitem[{{Howell} {et~al.}(2005)}]{2005ApJ...634.1190H} 
{Howell}, D.~A. {et~al.} 2005, \apj, 634, 1190 

\bibitem[{{Ivanov}, {Hamuy}, \&  {Pinto}(2000)}]{2000ApJ...542..588I} 
{Ivanov}, V.~D., {Hamuy}, M., \& {Pinto}, P.~A.  2000, \apj, 542, 588 

\bibitem[{{Jha}, {Riess}, \&  {Kirshner}(2007)}]{2007ApJ...659..122J} 
{Jha}, S., {Riess}, A.~G., \& {Kirshner}, R.~P.  2007, \apj, 659, 122 

\bibitem[{{Kim} {et~al.}(2004)}]{2004MNRAS.347..909K} 
{Kim}, A.~G., {Linder}, E.~V., {Miquel}, R., \& {Mostek}, N.  2004, \mnras, 
347,  909 

\bibitem[{{Lentz} {et~al.}(2000)}]{2000ApJ...530..966L} 
{Lentz}, E.~J., {Baron}, E., {Branch}, D., {Hauschildt}, P.~H., \& 
{Nugent},  P.~E.  2000, \apj, 530, 966 

\bibitem[{{Li} {et~al.}(2003)}]{2003PASP..115..453L} 
{Li}, W. {et~al.} 2003, \pasp, 115, 453 

\bibitem[{{Linder}(2006)}]{2006APh....26..102L} 
{Linder}, E.~V.  2006, Astroparticle Physics, 26, 102 

\bibitem[{{Mannucci}, {Della Valle}, \&  
{Panagia}(2006)}]{2006MNRAS.370..773M} 
{Mannucci}, F., {Della Valle}, M., \& {Panagia}, N.  2006, \mnras, 370, 773 

\bibitem[{{Mannucci} {et~al.}(2005)}]{2005A&A...433..807M} 
{Mannucci}, F., {della Valle}, M., {Panagia}, N., {Cappellaro}, E., 
{Cresci},  G., {Maiolino}, R., {Petrosian}, A., \& {Turatto}, M.  2005, 
\aap, 433, 807 

\bibitem[{{Perlmutter} {et~al.}(1999)}]{1999ApJ...517..565P} 
{Perlmutter}, S. {et~al.} 1999, \apj, 517, 565 

\bibitem[{{Perlmutter} {et~al.}(1997)}]{1997ApJ...483..565P} 
{Perlmutter}, S. {et~al.} 1997,  \apj, 483, 565 

\bibitem[{{Phillips}(1993)}]{1993ApJ...413L.105P} 
{Phillips}, M.~M.  1993, \apjl, 413, L105 

\bibitem[{{Prieto}, {Rest}, \&  {Suntzeff}(2006)}]{2006ApJ...647..501P} 
{Prieto}, J.~L., {Rest}, A., \& {Suntzeff}, N.~B.  2006, \apj, 647, 501 

\bibitem[{{Riess} {et~al.}(1998)}]{1998AJ....116.1009R} 
{Riess}, A.~G. {et~al.} 1998, \aj, 116, 1009 

\bibitem[{{Riess}, {Press}, \&  {Kirshner}(1996)}]{1996ApJ...473...88R} 
{Riess}, A.~G., {Press}, W.~H., \& {Kirshner}, R.~P.  1996, \apj, 473, 88 

\bibitem[{{Riess} {et~al.}(2007)}]{2007ApJ...659...98R} 
{Riess}, A.~G. {et~al.} 2007, \apj, 659, 98 

\bibitem[{{Scannapieco} \& {Bildsten}(2005)}]{2005ApJ...629L..85S} 
{Scannapieco}, E. \& {Bildsten}, L.  2005, \apjl, 629, L85 

\bibitem[{{Strolger} {et~al.}(2004)}]{2004ApJ...613..200S} 
{Strolger}, L.-G. {et~al.} 2004, \apj, 613, 200 

\bibitem[{{Sullivan} {et~al.}(2006)}]{2006ApJ...648..868S} 
{Sullivan}, M. {et~al.} 2006, \apj, 648, 868 

\bibitem[{{Timmes}, {Brown}, \&  {Truran}(2003)}]{2003ApJ...590L..83T} 
{Timmes}, F.~X., {Brown}, E.~F., \& {Truran}, J.~W.  2003, \apjl, 590, L83 

\bibitem[{{Tonry} {et~al.}(2003)}]{2003ApJ...594....1T} 
{Tonry}, J.~L. {et~al.} 2003, \apj, 594, 1 

\bibitem[{{Weller} \& {Albrecht}(2002)}]{2002PhRvD..65j3512W} 
{Weller}, J. \& {Albrecht}, A.  2002, \prd, 65, 103512 

\end{thebibliography}


\end{document}